\documentclass[review]{elsarticle}

\usepackage{lineno,hyperref}
\usepackage{amsmath}
\modulolinenumbers[5]
\usepackage{xparse}
\journal{NeuroImage}
\usepackage{lineno}
\usepackage{graphicx}
\usepackage{algorithm}
\usepackage{algpseudocode}
\usepackage[nobiblatex]{xurl}
\usepackage{diagbox}

\makeatletter

\begin{document}

\begin{frontmatter}

\title{ Discovering Dynamic Effective Connectome of Brain with Bayesian Dynamic DAG Learning}
\author[mymainaddress]{Abdolmahdi Bagheri\corref{mycorrespondingauthor}}
\cortext [mycorrespondingauthor] {Corresponding author}
\ead{Abdolmahdibagheri@ut.ac.ir}
\author[mymainaddress]{Mohammad Pasande}
\author[mythirdaddress]{Kevin Bello}
\author[mymainaddress]{Babak Nadjar Araabi}
\author[mysecondaryaddress]{Alireza Akhondi-Asl}

\address[mymainaddress]{School of Electrical and Computer Engineering, University of Tehran, College of Engineering, Tehran, Iran}
\address[mythirdaddress]{Machine Learning Department, Carnegie Mellon University, Pittsburgh, USA}
\address[mysecondaryaddress]{Department of Anaesthesia, Harvard Medical School, Boston, Massachusetts, USA}

\begin{abstract}
Understanding the complex mechanisms of the brain can be unraveled by extracting the Dynamic Effective Connectome (DEC).  Recently, score-based Directed  Acyclic Graph (DAG) discovery methods have shown significant improvements in extracting the causal structure and inferring effective connectivity. However, learning DEC through these methods still faces two main challenges: one with the fundamental impotence of high-dimensional dynamic DAG discovery methods and the other with the low quality of fMRI data.
In this paper, we introduce Bayesian Dynamic DAG learning with M-matrices Acyclicity characterization \textbf{(BDyMA)} method to address the challenges in discovering DEC. The presented dynamic causal model enables us to discover direct feedback loop edges as well.
Leveraging an unconstrained framework in the BDyMA method leads to more accurate results in detecting high-dimensional networks, achieving sparser outcomes, making it particularly suitable for extracting DEC. 
Additionally, the score function of the BDyMA method allows the incorporation of prior knowledge into the process of dynamic causal discovery which further enhances the accuracy of results. 
Comprehensive simulations on synthetic data and experiments on Human Connectome Project (HCP) data demonstrate that our method can handle both of the two main challenges, yielding more accurate and reliable DEC compared to state-of-the-art and traditional methods. Additionally, we investigate the trustworthiness of DTI data as prior knowledge for DEC discovery and show the improvements in DEC discovery when the DTI data is incorporated into the process.
\end{abstract}
\end{frontmatter}

\section{\textbf{Introduction}}
\label{sec:introduction}

{T}{he} studies on the brain's connectome are motivated by the goal of understanding the intricate interactive functionality of its regions. Studies such as \cite{gerstner2000population,beurle1956properties,wilson1973mathematical} have illustrated the transient patterns of coordinated activity that evolve at the brain neural level over time. Consequently, studies by Chang et al. \cite{Chang2010Time} and Preti et al. \cite{Preti2017dynamic} have revealed that the extraction of connectivity among different brain regions can also be more accurately modeled as dynamic networks.
The investigation of characterizing these dynamic networks' behavior is explored in \cite{Roberts2019Metastable, Varley2022Network, Vidaurre2017Brain, Vidaurre2018Brain, kam2019deep}, primarily focusing on employing statistical dependencies between regional activities, commonly referred to as Dynamic Functional Connectivity (DFC), to understand how the brain functions.
However, the Dynamic Effective Connectome (DEC) goes beyond DFC by exploring how one region influences another over time, indicating causal interactions among brain regions. 

One of the earliest attempts in modeling the causal dynamic behaviors of the brain was undertaken by Friston et al., who proposed the Dynamical Causal Modeling (DCM) in \cite{Friston2003Dynamic}. 
However, this method integrates varying levels of prior biological plausibility to comprehend the mechanisms behind neuroimaging data \cite{camps2023discovering}.
On the other hand, there are other methods that are developed to discover causality in the brain based solely on observational and interventional data. For instance, methods such as stimulation of specific regions \cite{siddiqi2021brain,siddiqi2022causal}, Granger causality \cite{roebroeck2005mapping,Bressler2011Wiener} or LiNGAM \cite{spielberg2015brain,ramsey2014non} focus on discovering causality within a small number of regions across the numerous areas in the Brain. However, the accuracy of these methods diminishes when confronted with high-dimensional networks. A more in-depth review on the topic of causality in the brain is studied in \cite{camps2023discovering}.
Thus, the main concern lies in the disparity between the limitations of providing low-dimensional causal statements and the demand for high-dimensional causal statements.

The search for alternative approaches of causal discovery that address the mentioned concern is of great interest in gaining a better understanding of brain mechanisms. The score-based approaches, particularly those employing continuous optimization problems for extracting Directed Acyclic Graphs (DAGs), are shown to yield higher accuracy in extracting the causal structure of high-dimensional networks. These approaches originated with the NOTEARS method presented in \cite{zheng2018dags}, which formulated the causal structure learning problem as a continuously constrained optimization task instead of relying on local heuristics for enforcing the acyclicity constraint. Building upon this approach, Yu et al. \cite{yu2019dag} altered the score function to extract nonlinear structures. In \cite{Geffner2022deep}, search for nonlinear structure advanced with proposing a new score function.
In a related study, Zhang et al. \cite{Zhang2022Detecting} employed the extended version of the NOTEARS method to detect abnormal connectivity in schizophrenia. In \cite{ng2020role}, the authors presented the GOLEM method,
which replaces the DAG hard constraint with a soft constraint in the score function and relies on thresholding to obtain a DAG solution and overcome the difficulties associated with constrained optimization. In \cite{bagheri2023brain}, the authors used the idea of the GOLEM method along with the employment of structural data as prior knowledge to extract effective connectome based on both fMRI and DTI data. 
In the following, Deng et al. \cite{deng2023optimizing} proposed a method based on topological swaps to improve the solutions of score-based methods using a continuous acyclicity formulation.
Additionally, Liu et al.  \cite{Liu2021Inferring} proposed a method that instead of DAGness constraints, combines the greedy equivalence search with a novel score function for inferring effective connectivity from fMRI data based on conditional entropy and transfer entropy between brain regions.

The NOTEARS idea is not limited to static causal structure extraction and has also been applied to extracting causal dynamic mechanisms. In the DYNOTEARS method \cite{Pamfil2020DYNOTEARS}, the authors used the NOTEARS idea to learn the underlying dynamic structure of a system. Similar to the NOTEARS method, DYNOTEARS extracts the causal structure of the system by minimizing the least square error objective as the score function in the optimization problem. In both of the methods, the square error objective is the distance between the model and the data with the same DAGness constraint in their optimization problem.
Dynamic DAG learning with continuous hard constraint optimization problems meets their application in extracting DEC as well. In \cite{Mansouri2020Interpretable}, the HEIDEGGER framework is proposed as an interpretable temporal causal discovery that has been tested on a cognitive health dataset to find out which lifestyle factors matter in reducing cognitive decline.
Yu et al. \cite{Yu2022Learning} proposed an end-to-end framework, TBDS, for fMRI analysis to better adapt GNNs for fMRI analysis. Moreover, in \cite{Liu2023Learning}, the NOTEARS-PFL method is proposed for simultaneous learning of multiple Bayesian networks with continuous optimization, to identify disease-induced alterations in major depressive disorder patients.
Yet, utilizing the task of discovering DEC from fMRI data commonly suffers from two main obstacles. Firstly, the accuracy of these methods decreases significantly with the size of the network \cite{Pamfil2020DYNOTEARS}. 
Second, measuring brain activity with the fMRI method is limited by the slow hemodynamic response, spatial and temporal resolution constraints, and susceptibility to confounding factors \cite{bennett2010reliable}. As a result, the fMRI cannot accurately capture brain activity due to its indirect measure of neural activity.

In this paper, we introduce a Bayesian causal framework, i.e., Bayesian Dynamic DAG learning with M-matrices Acyclicity characterization \textbf{(BDyMA)}, that addresses both of the mentioned challenges.
Our method enhances the effectiveness of dynamic DAG discovery for large-scale networks through two main steps. We formulate a likelihood-based score function with only the soft sparsity and DAG constraints to learn the ground truth DAG, which leads to an unconstrained optimization problem that is much easier to solve. This procedure enables extracting high-dimensional networks while retaining high accuracy. 
Using the Maximum Likelihood Estimator in the score function of the BDyMA method makes it more suited for extracting DEC based on fMRI data, where multivariate Gaussianity approximation is a reasonable assumption \cite{Sundaram2019Individual}.
Second, we use the DAGs via M-matrices for Acyclicity (DAGMA) \cite{Bello2022Learning} characteristic as the DAGness constraint, which is shown to be more effective in discovering DAGs compared to the one introduced in \cite{zheng2018dags}. The M-matrices are defined over the cone of positive definite matrices to deal with the asymmetries of a DAG which leads to a better detection of large cycles.
While these improvements significantly enhance the accuracy of the results in extracting dynamic causal mechanisms, extracting DEC is still severely constrained by the quality of fMRI data that leads to results with fundamental ambiguities such as low reliability \cite{maier2017challenge,sporns2013human}. 
As it is shown in \cite{eggeling2019structure}, the prior knowledge that favors sparsity (such as DTI \cite{tsai2018Reproducibility}) can significantly enhance both the accuracy and reliability of discovered graphs when a limited amount of data (such as fMRI signals) is available and the dimension of networks grows (such as brain networks). 
Furthermore, it is highly likely that the true DAG of high dimensional structure is non-identifiable \cite{cundy2021bcd}.
As a result, we address these issues by leveraging DTI data as prior knowledge in our approach which is shown to be effective in discovering DEC \cite{Xia2023Structure}.

The main contributions of our work are thus fourfold:
\begin{itemize}
    \item First, we introduce BDyMA as a Bayesian dynamic DAG framework to discover the dynamic causal structure of high-dimensional networks.
    \item Second, we demonstrate the effectiveness of the BDyMA in comparison to both the state-of-the-art method, DYNOTEARS \cite{Pamfil2020DYNOTEARS}, and the LiNGAM \cite{shimizu2011directlingam} in terms of accuracy and reliability. Additionally, using binary and probabilistic priors further advances the accuracy of the causal discovery.
    \item Third, we show that our Bayesian dynamic DAG discovery method enhances the intrasubject and intersubject {reliability} of DEC discovery.
    \item Fourth, we examine the trustworthiness of DTI data as prior knowledge of DEC discovery and highlight the potential for this framework to significantly advance our understanding of brain functionality.
\end{itemize}

In section \ref{method}, we present our proposed framework, BDyMA. Section \ref{results} illustrates 
the effectiveness of our methods by comparing the DEC discovered with BDyMA, DYNOTEARS, and LiNGAM methods. Then, we employ Bayesian causal frameworks with HCP data, assess the reliability of the discovered ECs through various means, and investigate the validity of DTI data as prior knowledge of DEC discovery. Finally, we discuss the results and limitations of our method. The codes associated with this study will be released upon the formal acceptance of the paper for publication.

\section{Method}\label{method}
In this section, we begin with introducing our Bayesian causal dynamic framework, \textbf{BDyMA}, and its mathematical backbone. Then, in the data subsection, we begin by discussing the Causal sufficiency assumption and its relation with our choice for parcellation, explaining its significance and implications for our study. Following that, we present the HCP data used in our study, providing details about the data acquisition and preprocessing procedures. Next, we describe the process of generating synthetic data and finally, we present the empirical data that we use in this paper.
\subsection{Dynamic DAG structure learning}
A dynamical causal model is defined by a distribution $p\bigl(X(t)\bigl)$, where $X(t)=[X_1(t),..., X_d(t)]^T $ is a stationary process, $d$ is the number of variables, $0<t\le n$ and $n$ is the number of observations in time series of each variable. 

\begin{figure}[!h]
\centering
\includegraphics[width=.5\textwidth]{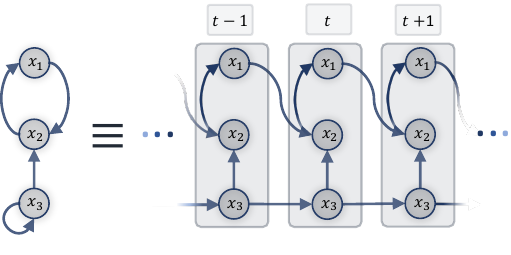}
\caption{An example of a cyclic dynamic structure with $m=1$}
\label{fig:1}
\end{figure}

We model a dynamic DAG structure with both instantaneous and time-lagged effects as follows
\begin{multline}
\label{eq:1}
{X}(t)=\boldsymbol{B^T}{X}(t)+\boldsymbol{A}_1{X}(t-1)+...
+ \boldsymbol{A}_m{X}(t-m)+ \epsilon(t)
\end{multline}
where $\epsilon(t)=\bigl(\epsilon_1(t), . . . , \epsilon_d(t)\bigl)$ is the noise vector that is a set of jointly Gaussian and independent exogenous variables. This vector has $N(0, \boldsymbol{\Omega})$ distribution, where the covariance matrix is $\boldsymbol{\Omega}=diag(\sigma_1^2,...,\sigma_d^2 )$. $\boldsymbol{B}$ is the causal structure of instantaneous effects that has to be a DAG and $\boldsymbol{A}_j, 0<j\le m$, shows the causal effect from the previous time slice to the current time slices. Contrary to the $\boldsymbol{B}$, there is no DAGness constraint on the $\boldsymbol{A}_j$ matrices. Note that $\boldsymbol{B,A}_j\in \boldsymbol{R}^{d\times d}$. Fig.\ref{fig:1} shows the graph of a dynamic DAG structure that can represent a cyclic structure.
As it is illustrated in this figure, this type of representation not only captures the dynamic behavior of a causal process but also represents a causal model with cycles. In other words, this framework is able to extract direct feedback loop edges within the resulting causal model. This type of modeling can address one of the key challenges associated with employing DAGs as a causal structure, as highlighted in studies like \cite{imbens2020potential}. 
\subsubsection{Optimization problem}
The following unconstrained optimization problem for finding the underlying dynamic DAG structure is designed based on the optimization problem in \cite{zheng2018dags} and is used to estimate the weighted adjacency matrices $\boldsymbol{B}$ and $\boldsymbol{A}_j$, given the data $X(t)$.
\begin{equation}
\begin{split}
\min_{\{\boldsymbol{B},\boldsymbol{A}_1,..., \boldsymbol{A}_m\}\in R^{d\times d} }  & S(\boldsymbol{B},\boldsymbol{A}_1,..., \boldsymbol{A}_m;X,\boldsymbol{G_P})  \\&
=L(\boldsymbol{B},\boldsymbol{A}_1,..., \boldsymbol{A}_m;X)  \\&
+R_{sparse}(\boldsymbol{B},\boldsymbol{G_P})+R_{sparse}(\boldsymbol{A},\boldsymbol{G_P})+\lambda_3h(\boldsymbol{B}) 
\label{eq:1_0}
\end{split}
\end{equation} 
where $S(.)$ is the score function, $\boldsymbol{G_P} \in R^{d\times d}$ is the prior knowledge (matrix) on the graph structure, 
$L(\boldsymbol{B},\boldsymbol{A}_1,...,\boldsymbol{ A}_m;X)$ is the Likelihood objective term, $\boldsymbol{A}=\bigl[\boldsymbol{A}_1| \boldsymbol{A}_2|, ..., |\boldsymbol{A}_m\bigl]$,  $\lambda_3$ is the penalty coefficients, and $h(\boldsymbol{B})$ can be any of the characterization of DAGness such as \cite{zheng2018dags,Bello2022Learning}.
The sparsity penalty terms in \ref{eq:1_0} include the prior probability as follows:

\begin{equation}
\begin{split}
&R_{sparse}(\boldsymbol{B},\boldsymbol{G_P})  ={\lambda_{\textrm{1}}} \Big\| Q \bigl(\boldsymbol{B},\boldsymbol{G_P}  \bigl)  \Big\| \\ &
R_{sparse}(\boldsymbol{A},\boldsymbol{G_P})={\lambda_{\textrm{2}}} \Big\| Q \bigl( \boldsymbol{A},\boldsymbol{G_P}  \bigl)  \Big\|
\label{eq:1_001}
\end{split}
\end{equation} 
where $Q(.)$ is an element-wise smooth monotonic function.
This function acts element-wise on the input matrices, such that if a specific element in the $\boldsymbol{G_P}$ matrix moves in one direction, the corresponding element of the $\boldsymbol{B}$ or $\boldsymbol{A}$ moves in the same direction. 
Such a sparsity penalty term inhibits/excites an element of the $\boldsymbol{B}$ or $\boldsymbol{A}$ matrices that has a lower/higher prior probability. In this paper, we have considered 
\begin{align}
\begin{split}
& Q \bigl( \boldsymbol{B},\boldsymbol{G_P}  \bigl)= \log\bigl({\boldsymbol{B}}\bigl) \odot {\boldsymbol{G_P}} \\ &
Q \bigl( \boldsymbol{A},\boldsymbol{G_P}  \bigl)= [\log\bigl({\boldsymbol{A_1}}\bigl) \odot {\boldsymbol{G_P}},...,\log\bigl({\boldsymbol{A_m}}\bigl) \odot {\boldsymbol{G_P}}]
\label{eq:7-5}
\end{split}
\end{align}
where $\odot$ is the element-wise matrix product.
The choice of the $Q(.)$ function can change the incorporation of prior knowledge in equation \ref{eq:7-5}.  Employing the $\log(.)$ function balances the skewness of distributions. In the context of incorporating prior knowledge into the final graph, the logarithmic transformation results in a more conservative approach. Moreover, the logarithmic transformation can help reduce the impact of noise on the data. 
For a binary prior knowledge, $Q \bigl( \boldsymbol{B},\boldsymbol{G_P} \bigl)$ and $Q \bigl( \boldsymbol{A},\boldsymbol{G_P} \bigl)$ are the element-wise product of a binary value to the adjacency matrices.

In this paper, we use the following $h(b)$ function, presented in \cite{Bello2022Learning} that does not have the drawbacks of other DAGness terms and is better at detecting large cycles, has better-behaved gradients, and is faster in practice.
\begin{equation}
\begin{split}
h(\boldsymbol{B}) =  \log \det{(s\boldsymbol{I}-\boldsymbol{B}\circ \boldsymbol{B})} - d\log s 
\label{eq:1_000}
\end{split}
\end{equation}
for $s>0$. $\circ$ is the Hadamard product.

In equation \ref{eq:1_0}, the likelihood objective in the score function is as follows
\begin{multline}
 L(\boldsymbol{B},\boldsymbol{A}_1 ,..., \boldsymbol{A}_m;X)=  
 \log{|\det(\boldsymbol{I-B})|}  \\
 -\frac{1}{2}\sum_{i=1}^d \left(\sum_{t=m+1}^n\left(X_i(t)-B_i^T(t){X}(t)-\boldsymbol{A}\boldsymbol{Y}(t)\right)^2\right)
 \label{eq1_1}
\end{multline}
where ${\boldsymbol{Y}(t)}=\bigl[X(t-1)^T,..., X(t-m)^T\bigl]^T$. 

For structures with equal noise variances, the likelihood function is
\begin{multline}
L(\boldsymbol{ B},\boldsymbol{A}_1 ,..., \boldsymbol{A}_m;X)=\log{|\det(\boldsymbol{I-B})|} 
\\
-\frac{d}{2}\log\sum_{i=1}^d\sum_{t=m+1}^n{{\left(X_i(t)-B_i^T{X}(t) 
-\boldsymbol{A_i}{\boldsymbol{Y}}(t)\right)^2}}
\label{eq:1_2}
\end{multline}

Instead of the likelihood as the objective function, one can use the square error value in the score function. However, as shown in \cite{wackerly2014mathematical}, the maximum likelihood estimation is an unbiased estimator, whereas the minimum square error estimation introduces bias.
It is worth noting that the primary objective of the optimization problem outlined in equation \ref{eq:1_0} is to find matrices $\boldsymbol{A}$ and {$\boldsymbol{B}$} that best fit the data, while ensuring that the $\boldsymbol{B}$ matrix forms a DAG. However, in this optimization problem, the covariance matrix is unknown, and one can profile out the $\boldsymbol{\Omega}$ function from the optimization problem by determining the extreme values of the likelihood function. This procedure is as follows.

\subsubsection{Deriving objective likelihood functions}
To derive the objective likelihood functions, from \ref{eq:1}, we have
\begin{multline}
\label{eq:2}
(\boldsymbol{I}-\boldsymbol{B^T})X(t)= \boldsymbol{A_1}{X}(t-1)+...+ \boldsymbol{A_m}{X}(t-m)+ \epsilon(t)
\end{multline}

Assuming that $(\boldsymbol{I} - \boldsymbol{B^T})$ is invertible, we rewrite the linear model in equation \ref{eq:2} as
\begin{multline}
\label{eq:3}
X(t)=\boldsymbol{(I-B^T)}^{-1}\boldsymbol{A}_1X(t-1)+...\\  + \boldsymbol{(I-B^T)}^{-1}\boldsymbol{A}_mX(t-m)+{Z}(t) 
= \boldsymbol{(I-B^T)}^{-1}\boldsymbol{AY}(t)+{Z}(t)
\end{multline}
where ${Z}(t) =  (\boldsymbol{I} - \boldsymbol{B^T})^{-1}\epsilon(t)$.
Then, the precision matrix of $\boldsymbol{Z}(t)$ is 
\begin{align}
\label{eq:5}
\boldsymbol{\Theta}=\boldsymbol{{\Sigma}^{-1}}=(\boldsymbol{I} - \boldsymbol{B})\boldsymbol{\Omega^{-1}}(\boldsymbol{I} - \boldsymbol{B})^T
\end{align} 

Due to the Markovian structure of the autoregressive  (AR) process, the joint density of the given data is 
\begin{equation}
\begin{split}
p\bigl(X(1)  ,... & ,X(n)|\boldsymbol{A}, \boldsymbol{B}, \boldsymbol{\Omega}\bigl) =\\&
p\bigl(X(1),...,X(m)|\boldsymbol{A}, \boldsymbol{B}, \boldsymbol{\Omega}\bigl)
\prod_{t=1+m}^np\bigl(X(t)|{\boldsymbol{Y}(t)}; \boldsymbol{A}, \boldsymbol{B}, \boldsymbol{\Omega}\bigl)
\label{eq:6}
\end{split}
\end{equation}

It is shown that finding the likelihood objective of probability density in equation \ref{eq:6} results in a nonlinear estimation problem \cite{kay1994maximum}. As a result, we derive the objective of the likelihood of conditional probability density as an alternative approximation of the main probability function conditioned
on the first $m$ observation.

The conditional probability of the data is
\begin{equation}
\begin{split}
p\bigl(X(m+1)  ,... & ,X(n)|X(1),..., X(m);\boldsymbol{A}, \boldsymbol{B}, \boldsymbol{\Omega}\bigl) = \prod_{t=1+m}^np\bigl(X(t)|{\boldsymbol{Y}(t)}; \boldsymbol{A}, \boldsymbol{B}, \boldsymbol{\Omega}\bigl)
\label{eq:7}
\end{split}
\end{equation}

From equation \ref{eq:3}, we have
\begin{align}
\label{eq:8}
X(t)-\boldsymbol{(I-B^T)}^{-1}\boldsymbol{A}{Y(t)}=Z(t)
\end{align} 
and as a result, from equation \ref{eq:7} and \ref{eq:8}, $p\bigl(X(t)|{Y(t)};\boldsymbol{A}, \boldsymbol{B}, \boldsymbol{\Omega}\bigl) \sim N\bigl(\boldsymbol{(I-B^T)}^{-1}\boldsymbol{A}{Y(t)}, \boldsymbol{\Theta^{-1}}\bigl)$ for $t>m$.

The log-likelihood function of the $AR$ model conditional on the $m$ initial values is 
\begin{multline}
\log p\bigl(X(m+1) ,...,X(n)|X(1),..., X(m);\boldsymbol{C}, \boldsymbol{\Theta}\bigl)=  \\
 -\frac{d(n-m)}{2}\log{(2\pi)}-\frac{n-m}{2}\log{\det(\boldsymbol{\Sigma})}  \\
-\frac{1}{2}\sum_{t=m+1}^n\bigl(X(t)-\boldsymbol{(I-B^T)}^{-1}\boldsymbol{A}\boldsymbol{Y}(t)\bigl)^T\boldsymbol{\Theta} \\ \bigl(X(t) -\boldsymbol{(I-B^T)}^{-1}\boldsymbol{A}\boldsymbol{Y}(t)\bigl)
\label{eq:9}
\end{multline}
Equation \ref{eq:9} can be rewritten as follows,
\begin{multline}
	\log  {p\bigl( X(m+1)  ,...,X(n)|X(1),..., X(m);\boldsymbol{C,B},\boldsymbol{\Omega}}\bigl) = \\
	-\frac{d(n-m)}{2}\log{(2\pi)} 
	+\frac{n-m}{2} {\log \det(\boldsymbol{I} - \boldsymbol{B})\boldsymbol{\Omega^{-1}}(\boldsymbol{I} - \boldsymbol{B})^T}  \\
	-\frac{1}{2}\sum_{t=m+1}^n\bigg(\bigl(X(t)-\boldsymbol{(I-B^T)}^{-1}\boldsymbol{A}\boldsymbol{Y}(t)\bigl)^T(\boldsymbol{I-B})  \\ \boldsymbol{\Omega^{-1}}(\boldsymbol{I-B})^T \bigl(X(t)-\boldsymbol{(I-B^T)}^{-1}\boldsymbol{A}\boldsymbol{Y}(t)\bigl)\bigg)=   \\
	const+(n-m)\log{|\det(\boldsymbol{I} - \boldsymbol{B})|} 
	-\frac{n-m}{2} \sum_{i=1}^d{\log \sigma^2_i}  \\
	-\frac{1}{2}\sum_{t=m+1}^n\sum_{i=1}^d{\frac{\bigl(X_i(t)-B_i^TX(t)-\boldsymbol{AY}{(t)}\bigl)^2}{\sigma^2_i}}
	\label{eq:10}
\end{multline}
and to profile out $\boldsymbol{\Omega}$, we have
\begin{multline}
	\hat{\sigma}_i^2(\boldsymbol{B,A}) = 
	\frac{1}{n-m}\sum_{t=m+1}^n{\bigl(X_i(t)-B_i^TX(t)  -\boldsymbol{A_iY}{(t)}\bigl)^2}.
	\label{eq:11}
\end{multline}
In this equation, the $\sigma^2_i$ values are profiled out with solving $\frac{\partial L}{\partial{\sigma^2_i}}=0$.
As a result, the objective likelihood is
\begin{multline}
L\bigl(\boldsymbol{B,A},\boldsymbol{\hat{\Omega}(\boldsymbol{B,A})};X \bigl)=  
 \log{|\det(\boldsymbol{I-B})|}  \\
 -\frac{1}{2}\sum_{i=1}^d \left(\sum_{t=m+1}^n\left(X_i(t)-B_i^T(t)X(t)-\boldsymbol{AY}(t)\right)^2\right).
\label{eq:12}
\end{multline}
Then, the goal is to find $\boldsymbol{B}$ and $\boldsymbol{A_j}$ that minimize the profile likelihood function.

If we consider the noise variances to be equal, i.e., $\sigma^2_i=\sigma^2$, the objective likelihood in equation \ref{eq:12}
yields to be
\begin{multline}
L\bigl(\boldsymbol{B,A},\boldsymbol{\hat{\Omega}(\boldsymbol{B,A})};X\bigl)  =
 const+(n-m)\log{|\det(\boldsymbol{I} - \boldsymbol{B})|}  -\frac{n-m}{2} \sum_{i=1}^d{\log \sigma^2_i}
\\-\frac{1}{2\sigma^2}\sum_{t=m+1}^n\sum_{i=1}^d{{\bigl(X_i(t) -B_i^TX(t)
-\boldsymbol{AY}{(t)}\bigl)^2}}
\label{eq:13}
\end{multline}
and with solving $\frac{\partial L}{\partial{\sigma^2}}=0$, the estimated $\sigma^2$ is
\begin{multline}
\hat{\sigma}^2(\boldsymbol{B,A}) = 
\frac{1}{n-m}\sum_{i=1}^d\sum_{t=m+1}^n{\bigl(X_i(t)-B_i^TX(t)
-\boldsymbol{AY}{(t)}\bigl)^2}
\label{eq:14}
\end{multline}
then, the objective likelihood is
\begin{multline}
L\bigl(\boldsymbol{B,A},\boldsymbol{\hat{\Omega}(\boldsymbol{B,A})};X\bigl)= \log{|\det(\boldsymbol{I-B})|}  \\
-\frac{d}{2}\log\sum_{i=1}^d\sum_{t=m+1}^n{{\bigl(X_i(t)-B_i^TX(t)
-\boldsymbol{AY}{(t)}\bigl)^2}}
\label{eq:15}
\end{multline}
\subsection{Algorithm}
Similar to the GOLEM method, the $\log\det(\boldsymbol{I-B})$ term is derived in the BDyMA method. While this term is zero if $\boldsymbol{B}$ corresponds to a DAG, the zero value of this term does not necessarily imply the DAGness of the $\boldsymbol{B}$ matrix \cite{Bello2022Learning}. As a result, similar to the DAGMA algorithm, algorithm \ref{alg:cap} is developed to discover the dynamic DAG structure and avoid the augmented Lagrangian scheme to tackle the problem.
Consequently, only soft sparsity and DAG constraints are necessary to learn a DAG equivalent to the ground truth DAG. 
\begin{algorithm}
 {\textbf{Input:}
  Data matrices ${X}$ and $ \boldsymbol{Y}$, initial coefficient $\mu^{(0)}$(e.g., 1), decay factor $\alpha \in (0, 1)$ (e.g., 0.1), $l_1$ parameter $\lambda_i > 0$ (e.g., 0.01), $s > 0$ (e.g., 1), number of iterations $T^\prime$.}
 \texttt{\\}
 \textbf{Initialization:} $\theta^{(0)}$ so that $\boldsymbol{B}(\theta^{(0)}) \in \boldsymbol{B}_s$ and $\boldsymbol{A}(\theta^{(0)})=0$.
 \begin{algorithmic}[1]
  \For{\texttt{$t = 0, 1, 2, \ldots, {T^\prime}{-1}$}}\\
  \textbf{Step $t$:}\\
  Starting at $\theta^{(t)}$, solve $\theta^{(t+1)} = \arg\min_{\theta} \mu^{(t)} \bigl(L(\boldsymbol{B}(\theta),\boldsymbol{A}(\theta),X,\boldsymbol{Y}) + \lambda_1 \lVert \boldsymbol{B}(\theta) \rVert_1+\lambda_2 \lVert \boldsymbol{A}(\theta) \rVert_1\bigl) + h(\boldsymbol{B}(\theta))$.\\
  Set $\mu^{(t+1)} = \alpha \mu^{(t)}$.
  \EndFor
 \end{algorithmic}
 {\textbf{Output:}
  $\boldsymbol{A}(\theta^{(T)}),\boldsymbol{B}(\theta^{(T)})$.}
 \caption{The BDyMA algorithm}\label{alg:cap}
\end{algorithm}

This algorithm returns a DAG whenever $\mu^{(t)} \xrightarrow{}  0$, solving the unconstrained optimization problem. The $\boldsymbol{B}_s$ represents the estimated $\boldsymbol{B}$ value based on the parameter $s$, which can take any value greater than zero. Importantly, it should be noted that any value of $s$ yields the same $\boldsymbol{B}$ matrix \cite{Bello2022Learning}.
To numerical solve the optimization problem, any gradient-based method can be employed in the algorithm. In this paper, we used the L-BFGS-B \cite{liu1989limited} due to its efficiency in large-scale batch optimization.
\subsection{Data}
\subsubsection{Synthetic data}
Synthetic data is produced utilizing the models elaborated in \cite{zheng2018dags}. This procedure entails the generation of three types of graphs by the node count to be $50, 75, 100, 125,$ and $150$, accompanied by 500 sample time points. 
The “ER2-ER2 with Gaussian noise" dataset is generated based on the Erdős–Rényi graphs, with both matrices $\boldsymbol{B}$ and $\boldsymbol{A}$ having a degree of 2, accompanied by additive Gaussian noise.
Similarly, the “ER2-ER2 with Non-Gaussian noise" dataset is generated in the same way, but with exponential noise.
In the third group of data, denoted as “BA4-ER2 with Gaussian noise" the $\boldsymbol{B}$ matrices are generated with the Barabasi–Albert model and degree of 4, while the $\boldsymbol{A}$ matrices follow the Erdős–Rényi structure with a degree of 2. In this dataset, the noise is additive Gaussian.
For each group of datasets and each node count, a total of 20 datasets have been generated. The generated data aims to demonstrate validation of the BDyMA method compared to others in DAG discovery across various types of structures and different real-world datasets. It's important to note that this does not imply that brain networks necessarily follow scale-free or Erdős–Rényi structures.

\subsubsection{Causal sufficiency and selected Parcellation}
\begin{figure}[!htb]
\centering
\includegraphics[width=\textwidth]{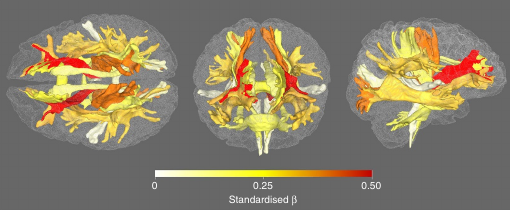}
\caption{Structural connectivities between two hemispheres, corpus callosum and subcortical regions\cite{cox2016ageing}.}
\label{fig:2_2}
\end{figure}

The causal sufficiency assumption seeks to answer the inquiry of whether additional variables should be taken into account during the causal discovery process \cite{spirtes2000causation}. 
In the brain, the subcortical regions along with the Corpus callosum are possible confounding factors for the cortical regions since they receive information flow through the stem. As is shown in Fig.\ref{fig:2_2}, there are no direct structural connections between the two hemispheres, and all the possible connections are mediated by the subcortical and Corpus callosum regions. 
As a result, these regions are included in our analysis to ensure that the causal sufficiency assumption is held. This means that there cannot be any direct effective connection between the two hemispheres, and all potential causal relationships between them are mediated through subcortical and Corpus callosum regions.
Correspondingly, we employ the Destrieux atlas \cite{destrieux2010automatic} that satisfies our concerns. The List of cortex regions is presented in Appendix 1.
It is worth noting that several other variables, such as sex, genetics, and heart rate, can act as confounding variables when studying brain effective connectivity. In the ideal scenario, these types of variables should be recorded alongside fMRI data and included in the set of variables for causal discovery.

\subsubsection{{Data acquisition}} In this study, we used DWI and {resting-state fMRI} data for 50 unrelated subjects that were provided by the “HCP 1200” data set \cite{van2013wu}. 
The HCP data sets were acquired using protocols approved by the Washington University institutional review board, and written informed consent was obtained from all subjects.
\subsubsection{{DTI data preprocessing}}
DWI images were acquired using a 3T ‘Connectome Skyra’, provided with a Siemens SC72 gradient coil and stronger gradient power supply with maximum gradient amplitude (Gmax) of 100 $mT/m$ sequence and 90 gradient directions equally distributed over 3 shells (b-values 1000, 2000, 3000 $mm/{s^2}$ ) with 1.25mm isotropic voxels \cite{sotiropoulos2013advances}.
All the preprocessing steps are performed on the data with the guidelines presented in \cite{Glasser2016the}.
In this study, we employed probabilistic tractography that considers multiple possible pathways at each voxel to find any possible structural connection. 
The MRtrix3 software is used to perform whole-brain probabilistic tractography. The tckgen algorithm is then used to generate 50 million streamlines from then Fiber Orientation Distribution (FOD) images with a maximum tract length of 250 mm to obtain whole-brain probabilistic tractography. Finally, the tck2connectome function is employed to find the number of streamlines that started from each ROI and ended in all other ROIs to determine the structural connectivity between all nodes.

\subsubsection{Empirical data}
The HCP MRI data acquisition protocols and procedures have previously been described in full detail \cite{smith2013resting}.
We used the minimally preprocessed images of {resting state fMRI}, `rsfMRI'. The fMRI resting-state runs were acquired in four separate sessions on two different days, with two different acquisitions (left to right or LR, and right to left, or RL) per day \cite{glasser2013minimal}. 
To eliminate any potential systematic bias or artifacts that may occur due to the acquisition process, we average the time series data of two sessions for each day \cite{Physiological2011Triantafyllou}. In our study, the mean of the first day's two sessions is used for each subject as test data (Mean 1), while the mean of two sessions on the second day is used as retest data (Mean 2).

\section{Result}\label{results}
In this section, we demonstrate the effectiveness of our method by applying it to both synthetic and empirical data. Through this investigation, we aim to show how our approach enhances the accuracy and reliability of the results. Furthermore, we highlight the significant improvements achieved by utilizing binary and probabilistic priors in terms of accuracy and reliability.
Last but not least, we present evidence supporting the use of DTI data as valid prior knowledge for extracting DEC. 

\subsection{Results for the synthetic data}
We applied our method with and without leveraging 
prior knowledge to the synthetic data and compared the results with the DYNOTEARS and LiNGAM methods which represent the state-of-the-art and more traditional approaches, respectively. Subsequently, we examined the impact of prior knowledge on the accuracy of the results. For this purpose, we compared the outcomes of three scenarios: (1) BDyMA without prior knowledge, (2) BDyMA with binary prior, and (3) BDyMA with probabilistic prior.

\begin{figure}[!htb]
\centering
\includegraphics[width=.45\textwidth]{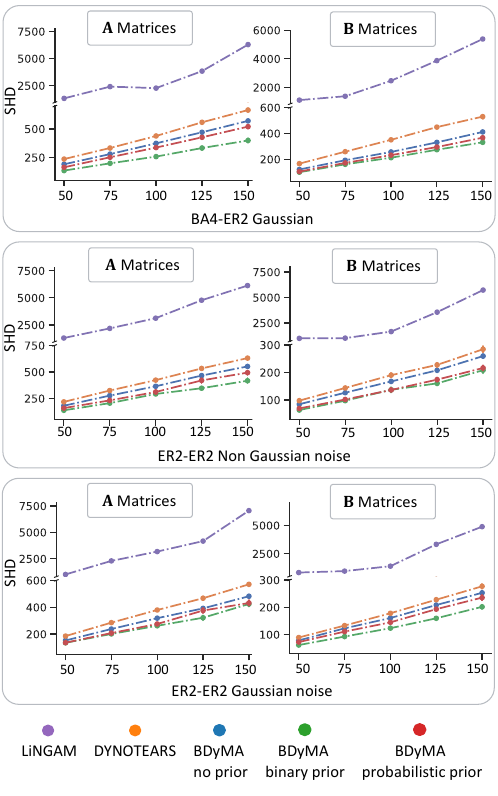}
\caption{SHD values of $\boldsymbol{A}$ and $\boldsymbol{B}$ matrices for data 1, data 2, and data 3 extracted with the BDyMA without prior, with binary prior knowledge and probabilistic prior knowledge, LiNGAM, DYNOTEARS, Number of Nodes $ \in \{50,75, 100,125,150\}$}
\label{fig:2_0}
\end{figure}
In Fig.\ref{fig:2_0}, we present a comparison of the BDyMA method's performance without prior knowledge, with binary prior knowledge, and with probabilistic prior knowledge against the DYNOTEARS and LiNGAM methods. In this figure, Structural Hamming Distance (SHD) values for the $\textbf{B}$ and $\textbf{A}$ of the three methods are computed. The SHD is the sum of the number of edge insertions, deletions or flips to transform one graph to the ground truth graph.
It is worth noting that we utilized thresholds of $0.3$ and $0.1$ to binarize the $\textbf{B}$ and $\textbf{A}$ matrices, respectively, across all datasets and methods.
The study in \cite{Pamfil2020DYNOTEARS} demonstrates the superior accuracy of the DYNOTEARS method compared to all other existing causal discovery methods in the literature. As a result, we compare our results with the DYNOTEARS method. This figure clearly indicates that the BDyMA method surpasses the accuracy of the DYNOTEARS approach by a significant margin.
Enhancing the accuracy of our method with the incorporation of binary and probabilistic prior knowledge narrows down the gap between the extracted adjacency matrix with the true structure.
The results of employing prior knowledge in the BDyMA may vary depending on the trustworthiness of the available prior knowledge and how much we incorporate it into our approach. This incorporation can be adjusted by changing the binarization threshold and the $Q(.)$ function in equation \ref{eq:7-5}.

\subsection{Results for the empirical data}
The primary step in the DEC extraction process involves determining the appropriate value of the autoregressive order, $m$. Following the previous studies in the literature such as \cite{li2008dynamic,rajapakse2007learning}, we adopt the assumption that $m=1$. 
Next, to assess the effectiveness of our proposed method on empirical data, we apply our method without incorporating prior knowledge and DYNOTEARS into the dataset. Then, we compare the results in terms of intersubject and intrasubject reliabilities. Subsequently, we discuss the comparison of derived DECs with structural and functional connectomes. Then, we investigate the impact of incorporating DTI data as the prior knowledge of DEC discovery.


\begin{figure*}[hbt!]
\centering
\includegraphics[width=1\textwidth]{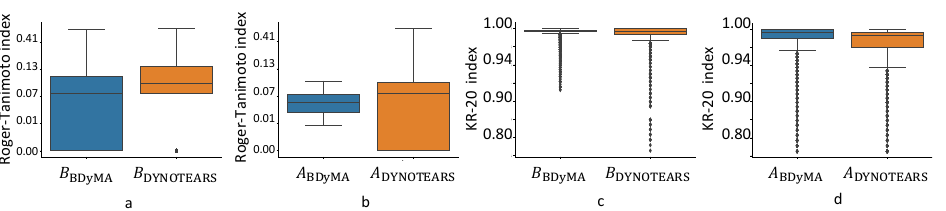}
\caption{The Rogers-Tanimoto and KR-20 values for all the edges with the BDyMA and DYNOTEARS method. a. The Rogers-Tanimoto index for DECs of the $\boldsymbol{B}$ matrices. b. The Rogers-Tanimoto index for DECs of the $\boldsymbol{A}$ matrices. C. The KR-20 index for DECs of the $\boldsymbol{B}$ matrices. D. The KR-20 index for DECs of the $\boldsymbol{A}$ matrices. }
\label{fig:3}
\end{figure*}

\subsubsection{Reliability tests} To assess the effectiveness of the BDyMA on the {reliability} of discovered DECs, we utilized the Rogers-Tanimoto index \cite{rogers1960computer} and Kuder-Richardson Formula 20 (KR-20) index \cite{Kuder1937theory} to compute intrasubject and intersubject reliability, respectively. The Rogers-Tanimoto index calculates dissimilarity between the edges of DECs obtained from the test and retest data of 63 subjects. The resulting index value for each edge ranges from 0 to 1, where 0 indicates perfect agreement and no dissimilarity of an edge in two sets of derived ECs, while 1 denotes no agreement and complete dissimilarity between an edge. Similarly, the KR-20 provides a reliability coefficient that ranges from 0 to 1. However, in this metric, higher values indicate greater reliability or agreement of an edge among the 63 subjects (intersubject).

In Fig.\ref{fig:3}, we demonstrate the Rogers-Tanimoto and KR-20 values for all the edges of the ECs with 164 regions extracted using the BDyMA and DYNOTEARS methods.
The median and interquartile range of the Rogers-Tanimoto index for DECs of the $\boldsymbol{B}$ matrices extracted with the BDyMA method are $0.071$ and $[0, 0.095]$, respectively. These values for the DECs of the $\boldsymbol{A}$ matrices are $0.051$ and $[0.032, 0.074]$. The median and interquartile range for the $\boldsymbol{B}$ matrices of the DYNOTEARS method are $0.11$ and $[0.071, 0.132]$, and for the $\boldsymbol{A}$ matrices, these values are $0.073$ and $[0, 0.09]$. 
The median and interquartile range of the  KR-20 index for DECs extracted with the BDyMA method are $0.991$ and $[0.988, 0.997]$ for the $\boldsymbol{B}$ matrices. The results for $\boldsymbol{A}$ matrices of this method are $0.989$ and $[0.979, 1]$. These values for the $\boldsymbol{B}$ matrices  of the DYNOTEARS method are $0.985$ and $[0.983, 1]$. The results for $\boldsymbol{A}$ matrices of this method are $0.975$ and $[0.96, 0.98]$. 
The significance of reliability improvement with our methods is tested with the Wilcoxon signed rank test.
The p-values of this test obtained for comparing the Roger-Tanimoto and KR-20 values of $\boldsymbol{B}$ and $\boldsymbol{A}$ matrices of two methods are extremely small, with a value lower than 1e{-3}. As a result, the reliability improvements are significant when BDyMA is employed.

\subsubsection{Comparison of sparsity} 
\begin{figure}[!h]
\centering
\includegraphics[width=.45\textwidth]{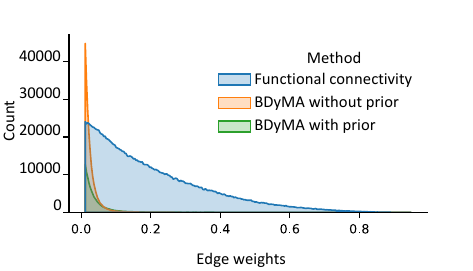}
\caption{ Weight density distributions of edges in functional connectivities and $\boldsymbol{B}$ matrices extracted with and without prior knowledge}
\label{fig:7_1}
\end{figure}

To deepen our investigation on the sparsity of the edges of DECs discovered with and without prior knowledge, Fig.\ref{fig:7_1} illustrates the weights of edges that are extracted in $63$ subjects and compares them to the edges of functional connectivities extracted based on correlation. For further clarity, we exclude edges with weights less than $0.01$.
According to Fig.\ref{fig:7_1}, the enclosed areas of the curves corresponding to the extracted DECs using our method are significantly lower than the functional connectivities. This discrepancy suggests a greater sparsity in the DECs. Furthermore, the incorporation of the prior values leads to a reduction in the area under the curve and higher weights, emphasizing the significance of utilizing DTI as a form of prior knowledge for DEC discovery.

\subsubsection{Comparison of connectomes} 
\begin{figure*}[!htb]
\centering
\includegraphics[width=\textwidth]{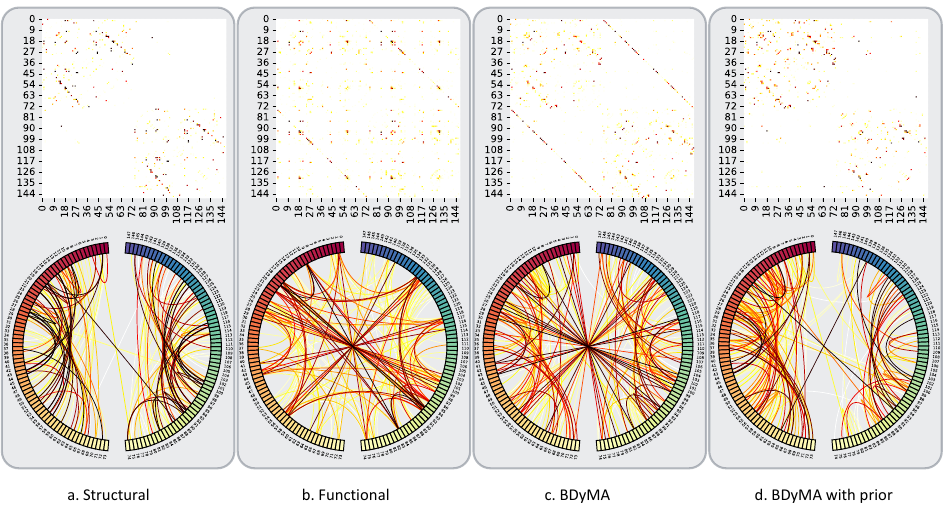}
\caption{Comparison of the Structural connectome, functional connectome, and the $\boldsymbol{B}$ matrices of the DECs discovered with the BDyMA method with and without incorporation of binary prior. The first row compares the symmetry of structural connectomes, correlation-based, and DECs and edges between the two hemispheres of each connectome. The second row compares the results of DECs discovered with $\textrm{SC}_\textrm{B}$, functional connectome, and  DECs of the BDyMA method with and without incorporation of binary prior  }
\label{fig:6}
\end{figure*}

Fig.\ref{fig:6} visualizes the structural connectome, the functional connectome, and the $\boldsymbol{B}$ matrices of the DECs extracted with the BDyMA, with and without prior knowledge. For enhanced visualization and comparative analysis of direct connections between the two hemispheres, only the regions in the two hemispheres are included in this figure.
Each plot in this figure is derived from the average of binarized connectomes with 250 edges.
The edges with warmer colors are those that appear in the majority of subjects, while cooler colors indicate less prevalent edges.

\textbf{Structural connectome:} Fig.\ref{fig:6}.a demonstrates that the structural connections within each hemisphere are strong. On the other hand, there are almost no direct structural connections between the two hemispheres, as it is elaborated in Fig.\ref{fig:2_2}.

\textbf{Functional connectome:} The functional connectivity illustrated in Fig.\ref{fig:6}.b is dominated by weak connections, exhibiting a higher degree of randomness in connectomes among subjects. Of particular note are the strong connections within hemispheres observed between the regions in the Insula and Intraparietal Sulcus in the right hemisphere and the regions in the Insula and Intraparietal Sulcus in the left hemisphere. Moreover, there are strong connections between hemispheres, e.g., the connections between regions in right Insula and left Insula, right Intraparietal Sulcus and left Intraparietal Sulcus, right Insula and left Intraparietal Sulcus, and left Insula and right Insula.
The connections between the right Insula and left Intraparietal Sulcus are also discovered in effective connectomes. Additionally, the connections between the left Insula and left Intraparietal Sulcus are discovered in both structural and effective connectomes. As a result, the triangle created by regions in the right insula and left Intraparietal Sulcus and left insula is likely generated due to the mediation of the left Intraparietal Sulcus. The same goes for the left Insula, right Intraparietal Sulcus, and right Insula.
It's worth mentioning that the regions in the right insula regions are in visual processing, while regions in the left intraparietal sulcus are known for their involvement in a variety of functions, including interoception, emotion processing, and social cognition. Such connections between brain regions may emerge based on specific cognitive processes or tasks under investigation. However, in studies examining the rest data, such as ours, these regions might not appear to have a direct effective relationship. This type of connectivity explains why functional connectivity can potentially yield misleading conclusions in causal inference. As expected, these kinds of connections are absent in the effective connectivities displayed in Fig.\ref{fig:6}.c and d, as well as in the structural connectivities. 

\textbf{Effective connectome:} Fig.\ref{fig:6}.c and d show that the $\boldsymbol{B}$ matrices of the DEC discovered with and without prior knowledge are dominated by strong connections, showing a higher degree of robustness in discovering DEC among subjects with our method contrary to the functional connectivities. The $\boldsymbol{B}$ matrices in Fig.\ref{fig:6}.c, reveal that all the direct effective connections between hemispheres are mostly along two diagonals parallel to the main diagonal. This implies that there should be direct causal relationships between regions in one hemisphere and their symmetrical counterparts in the other hemisphere. However, distinct tasks are undertaken by regions of each hemisphere meaning that the functionality of a region in one hemisphere does not necessarily directly affect the activity of the symmetrical region in the opposing hemisphere, as discussed in \cite{gilson2016estimation, toga2003mapping, davidson1995brain, kimura1973asymmetry,dolcos2002hemispheric}.
Moreover, as it is discussed on structural connectomes, such effective connections cannot exist due to the lack of any structural connections.
Consequently, the high correlation observed in symmetric regions should be either a result of confounding variables within the corpus callosum or subcortical regions or be attributed to spurious connections. To address this, acknowledging the fulfillment of the causal sufficiency assumption, all the effective paths between two hemispheres should be mediated by the regions in the corpus callosum or subcortical regions. 
As a result, the main question is to find the reasons for the discovery of such direct connections in studies, including our own. According to our review of the recent studies in discovering effective connectome, a prevalent trend emerges: these studies tend to neglect the causal sufficiency assumption by omitting potential confounding variables and analyzing only the cortex regions or smaller networks of the brain. This oversight can introduce confounding bias into the resulting graph, undermining the validity of the drawn causal inferences.
To address this question on the result of the BDyMA method, it is essential to delve deeper into the distinctions between Directed Acyclic Graphs (DAGs) and causal graphs.
While studies such as \cite{zheng2018dags, Bello2022Learning} showed higher accuracy of methods with DAGness constraint compared to other existing methods, it is important to note that these methods do not extract the causal graph itself. Instead, they estimate a DAG that best fits the observed data.
As a result, due to the high correlation between the signals of brain symmetric regions, drawing causal links between these symmetric regions increases the likelihood value without violating the DAGness constraint. Consequently, these methods identify such graphs as the best candidates for the true causal mechanism.
However, in light of structural connectivity and the causal sufficiency assumption, these edges are most likely to be false positives which highlights the importance of leveraging structural data in causal discovery methods. The incorporation of binary prior knowledge eliminates these types of connections and the DEC discovery is achieved by enforcing the estimation on the remaining edges. This ensures that the discovered DEC in Fig.\ref{fig:6}.d can be more accurate than the ones illustrated in  Fig.\ref{fig:6}.c.

\subsubsection{$\boldsymbol{A}$ matrices and the possible bidirectional edges} 
\begin{figure}[!htb]
\centering
\includegraphics[width=\textwidth]{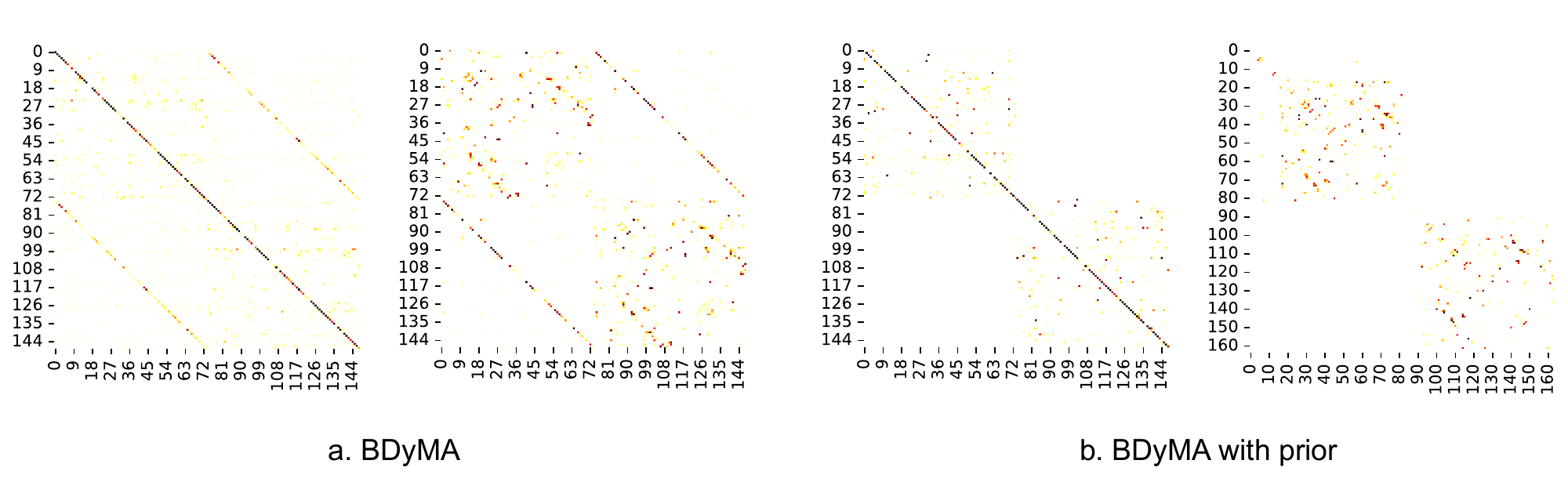}
\caption{ Comparison of the effective connections of the $\boldsymbol{A}$ matrices of the DECs discovered with the BDyMA method with and without incorporation of binary prior. }
\label{fig:7}
\end{figure}

Fig.\ref{fig:7} compares the $\boldsymbol{A}$ matrices of the DECs in scenarios with and without incorporation of binary prior knowledge.
The strong diagonal elements within the $\boldsymbol{A}$ matrices highlight the significant effect of all the regions from their value in the previous time slice to the current time. Similar to Fig.\ref{fig:6}.c, Fig.\ref{fig:7} displays diagonals that are parallel to the main diagonal, offering insights into the effect of these regions from the previous time. Additionally, for each of the A matrices, we have presumed that a node is not influenced by its prior value. Under this assumption, we observe more significant effects from one node to another across consecutive time points. It is essential to delve deeper into a comparative analysis of these matrices against those that permit influences from previous to current time points, with particular emphasis on the principles of Granger causality.  
As it is discussed on Fig.\ref{fig:6}.c, the parallel diagonals are most likely to be false positives as well. As a result, the same prior knowledge is employed for discovering $\boldsymbol{A}$ matrices as shown in Fig.\ref{fig:7}.b.
The edges that are in both  $\boldsymbol{A}$ and $\boldsymbol{B}$ matrices can signify direct feedback loop connections (as it is elaborated in Fig.\ref{fig:2_0}) when we want to model brain effective connectome statically. This addresses a limitation of causal discovery approaches employing the DAGness constraints, as discussed in \cite{Ji2021Estimating}.
Notably, the elements of $\boldsymbol{A}$ that do not find a counterpart in $\boldsymbol{B}$ highlight regions that are only affected from the previous time. This interplay between $\boldsymbol{A}$ and $\boldsymbol{B}$ matrices, along with direct feedback loop modeling within the DEC, provide a comprehensive reason for the necessity of dynamic causal modeling of brain mechanisms.

\subsubsection{Validity of DTI data as prior knowledge in DEC discovery} To further evaluate the validity of DTI data as prior knowledge in the discovery of DECs, we compare the reliability of discovered DECs with two different binarized prior matrices extracted from the DTI data.
In the first prior matrix, true binary SC denoted as SC$_{TB}$, we binarize the SC$_P$ matrix, assigning a value of $1$ to edges with higher values and $0$ to edges with lower values. In the second prior matrix, false binary SC denoted as SC$_{MB}$, the binarization process is reversed. We assign a value of $1$ to the edges with the lowest values in SC$_P$, and a value of $0$ to edges with higher values in SC$_P$, implying their lower relevance. Table. \ref{table:4} compares the score and reliability values for DEC discovered with the BDyMA method without prior knowledge and with truthful, SC$_{TB}$, and misleading, SC$_{MB}$, prior knowledge.

\begin{table}[]
\centering
\caption{The scores and reliability values of DEC discovered with BDyMA method without prior knowledge and with truthful, SC$_{TB}$, and misleading, SC$_{MB}$, prior knowledge}
\label{table:4}
\begin{tabular}{l|lll}\hline
\diagbox[width=.2\textwidth]{Method}{Metric}&
     {Score} &  KR-20 & Rogers-Tanimoto  \\
  \hline
No prior        & $453 \pm 20 $ & $[0.981,1] $  & $[0,0.082]$ \\
with SC$_{TB}$ & $461 \pm 21 $ & $[0.991,1]$ & $[0,0.043]$ \\
with SC$_{MB}$ &  $465 \pm 22$ & $[0.982,1]$  & $[0,0.066]$ \\
\hline
\end{tabular}
\end{table}
\subsubsection{ Runtime for the BDyMA method}
One of the main challenges in extracting DAGs is the computational costs. The running time of the BDyMA method on each subject is less than 15 seconds on an INTEL CORE i7 12700k processor. The primary reason for such a small running time is twofold. Firstly, in the numerical solution of the optimization function, we have derived the gradient of the likelihood function in the closed mathematical form and employed it instead of relying on the CPU to derive it numerically. Secondly, the DAGness term we have utilized. As demonstrated in \cite{Bello2022Learning}, employing the DAGMA term to enforce DAGness in the graph requires minimal computational power. This is achieved as computing ${h}$ in equation \ref{eq:1_000} and $\nabla$ are empirically faster.
\section*{ Limitations and promising aspects of future research}
In this paper, we have shown the improvement of the discovered DECs by employing our framework on both synthetic and real-world data. 
However, given the various types of errors reported in this study, more accurate methods are necessary to effectively investigate why and how the brain performs.
Additionally, the Markovian choice is set to be 1 in our study which is used in all of the similar studies. We recognize the necessity for a more comprehensive study to investigate the Markovian choice, as we believe this parameter can significantly impact the results. However, given that our study did not focus on this aspect and the absence of a ground truth for determining the optimal Markovian choice, we did not delve deeper into it. Our attempts to compare optimization problem scores and identify the minimum value by varying the choice of Markovian yielded consistently increasing error values. We attribute this primarily to model error rather than solely the Markovian choice being 1. As a result, we opted for the value traditionally used in the literature, while acknowledging that your mentioned concern warrants further investigation in future studies. In the end, as it is mentioned, the causal sufficiency assumption plays a vital role in discovering a causal mechanism. It would be insightful to delve deeper into the investigation of the role of this assumption in studying brain networks.
\section{Conclusion}
In this paper, we have developed the BDyMA method to address the existing challenges in discovering dynamic causal structures focusing on the practical limitations of dynamic effective connectome discovery.
Leveraging an unconstrained optimization problem in the BDyMA method leads to more accurate results in detecting high-dimensional networks, achieving sparser outcomes, and significantly improving the runtime, making it particularly suitable for extracting DEC. Additionally, the score function of the optimization problem of the BDyMA method allows the incorporation of prior knowledge in DEC discovery. 
Our simulations on synthetic data and experiments on Human Connectome Project (HCP) data demonstrate higher accuracy and reliability of DECs compared to state-of-the-art and baseline methods. 
The results of the empirical data showed the importance of employing DTI data in our method to reach the accurate DEC of the brain.

\bibliography{mybibfile}

\end{document}